\documentclass[preprint,aps,prmaterials,citeautoscript,
,floats,byrevtex,floatfix,showpacs]{revtex4-1} 
\usepackage{bm}
\usepackage{amsmath}
\usepackage{amsfonts}
\usepackage{amssymb}
\usepackage{graphicx}
\usepackage{setspace}
\usepackage{booktabs} 
\usepackage{natbib}
\usepackage{xcolor}
\usepackage{float}

\begin{document}

\title{Origins of the hydrogen signal in atom probe tomography}

\author{Su-Hyun \surname{Yoo}}
\email{yoo@mpie.de}
\affiliation{Department of Computational Materials Design, Max-Planck-Institut f\"{u}r Eisenforschung GmbH, Max-Planck-Str. 1, D-40237 D\"{u}sseldorf, Germany}
\author{Se-Ho \surname{Kim}}
\affiliation{Department of Microstructure Physics and Alloy Design, Max-Planck-Institut f\"{u}r Eisenforschung GmbH, Max-Planck-Str. 1, D-40237 D\"{u}sseldorf, Germany}
\author{Eric \surname{Woods}}
\affiliation{Department of Microstructure Physics and Alloy Design, Max-Planck-Institut f\"{u}r Eisenforschung GmbH, Max-Planck-Str. 1, D-40237 D\"{u}sseldorf, Germany}
\author{Baptiste \surname{Gault}}
\email{b.gault@mpie.de}
\affiliation{Department of Microstructure Physics and Alloy Design, Max-Planck-Institut f\"{u}r Eisenforschung GmbH, Max-Planck-Str. 1, D-40237 D\"{u}sseldorf, Germany}
\affiliation{Department of Materials, Royal School of Mines, Imperial College, London, SW7 2AZ, United Kingdom}
\author{Mira \surname{Todorova}}
\affiliation{Department of Computational Materials Design, Max-Planck-Institut f\"{u}r Eisenforschung GmbH, Max-Planck-Str. 1, D-40237 D\"{u}sseldorf, Germany}
\author{J\"{o}rg \surname{Neugebauer}}
\affiliation{Department of Computational Materials Design, Max-Planck-Institut f\"{u}r Eisenforschung GmbH, Max-Planck-Str. 1, D-40237 D\"{u}sseldorf, Germany}

\date{\today}

\begin{abstract}

Atom Probe Tomography (APT) analysis is being actively used to provide near-atomic-scale information on the composition of complex materials in three-dimensions. In recent years, there has been a surge of interest in the technique to investigate the distribution of hydrogen in metals. However, the presence of hydrogen in the analysis of almost all specimens from nearly all material systems has caused numerous debates as to its origins and impact on the quantitativeness of the measurement. It is often perceived that most H arises from residual gas ionization, therefore affecting primarily materials with a relatively low evaporation field. In this work, we perform systematic investigations to identify the origin of H residuals in APT experiments by combining density-functional theory (DFT) calculations and APT measurements on an alkali and a noble metal, namely Na and Pt, respectively. We report that no H residual is found in Na metal samples, but in Pt, which has a higher evaporation field, a relatively high signal of H is detected. These results contradict the hypothesis of the H signal being due to direct ionization of residual H$_2$ without much interaction with the specimen's surface. Based on DFT, we demonstrate that alkali metals are thermodynamically less likely to be subject to H contamination under APT-operating conditions compared to transition or noble metals. These insights indicate that the detected H-signal is not only from ionization of residual gaseous H$_2$ alone, but is strongly influenced by material-specific physical properties. The origin of H residuals is elucidated by considering different conditions encountered during APT experiments, specifically, specimen-preparation, transportation, and APT-operating conditions by taking thermodynamic and kinetic aspects into account.

\end{abstract}

\maketitle

\section{\label{Sec: Introduction} Introduction}

Understanding the effect that the lightest and smallest atom, hydrogen (H), has on the physical properties of materials is of paramount importance. For instance, H triggers changes in the mechanical properties of metallic materials, such as a sudden and unpredictable loss of ductility and toughness, which is commonly referred to as hydrogen embrittlement~\cite{Johnson1875,Daw1983}. H also plays for instance a key role in modifying electronic properties in semiconductors~\cite{Stavola1988,Walle2006}. Despite its importance, the direct imaging of H has remained extremely challenging thereby limiting our understanding of its influence on materials.

Atom probe tomography (APT) has the ability to detect all elements irrespective of their mass~\cite{Gault2021}, and provides three-dimensional compositional mapping with sub-nanometer resolution~\cite{DeGeuser2020}. This unique combination of high spatial resolution and chemical sensitivity is necessary to enable observations and quantification of hydrogen at specific microstructural features within complex materials. In recent years, these abilities have triggered a surge of interest in the use of the technique to study hydrogen~\cite{Chen2017,Chen2020,BREEN2020108,Mouton2021}.

Interestingly in APT measurements, it has been known that characteristic peaks at 1, 2, and 3\,Da, corresponding to H$^+$, H$_2^+$, and H$_3^+$ species respectively, are always produced under high-fields at the surface of many metals. This was studied in detail by Tsong and co-workers in the 1980s~\cite{Tsong1983} who introduced low pressures of H$_2$ inside of the vacuum chamber of the atom probe. Using more modern instrument setups, similar observations have been reported for metals~\cite{Chang2019}, semiconductors~\cite{Tweddle2019, Rigutti2021} and insulators~\cite{Lu2017} with a signal originating either from residual gases from the chamber or H$_2$ from the specimen itself. The H-related peaks can be minimized by reducing the hydrogen content by heat treatment in vacuum~\cite{BREEN2020108}, or by modifying the surface of the specimen by oxidation of the deposition of H-barrier thin films (e.g., TiN~\cite{doi:10.1063/1.1597376}). 

The study of hydrogen in materials by APT has hence often involved isotope labeling, i.e., using deuterium instead of hydrogen, in order to facilitate identification of the trapping sites for hydrogen in the microstructure~\cite{Gemma2012}, with an emphasis on steels~\cite{Takahashi2012,Chen2017}.

In spite of a consensual perspective that H-related species detected by APT are unavoidable, there has been a long-standing debate regarding the origin of detected H atoms. Gaseous H$_2$ molecules are present within the analysis vacuum chamber even in extremely low pressure and temperature (e.g., 10$^{-14}$\,bar and 90\,K). These H$_2$ molecules can be ionized during the measurement either under the effect of only the electric field or, possibly, in combination with the laser pulse, and dissociate, leading to the detection of atomic H$^+$ ions. This detected H is considered as noise in the analysis of the mass spectrum, having nothing to do with the actual distribution of the hydrogen within the microstructure of the sample.

Kolli hypothesized controlling the relative amplitude of hydrogen peaks in mass spectra that the hydrogen originates only from the residual hydrogen~\cite{Kolli2017}. In contrast, Breen et al. proposed that a substantial fraction of the detected H was inside the specimen itself~\cite{BREEN2020108}, in line with observations by Chang et al.~\cite{Chang2019}. There are important differences between these two mechanisms. For instance, in the former case, H is initially in the form of gaseous H$_2$ that becomes ionized away from the specimen's surface. In the latter case, hydrogen can be already in its atomic form inside the material or chemisorbed on the surface, and must be desorbed and ionized from the specimen surface itself, potentially following surface diffusion. 

This uncertainty about the origin limits our ability to precisely quantify H concentration in materials by using APT. It is therefore necessary to enhance our fundamental understanding of the origin and behavior of H in APT in order to elucidate numerous open questions regarding
H-involving mechanisms in physics, chemistry, and materials science, including hydrogen trapping or grain boundary segregation of H in the context of hydrogen embrittlement.

From a theoretical perspective, the high reactivity of H and its strong impact on the electronic structure of materials hinders theoretical investigations with approximate methods, such as interatomic potentials. The use of first-principle calculations has enabled the significant progress achieved in theoretical understanding of materials with H within the last 25 years~\cite{Walle2006,Ozolins2009,Takahashi2012}. Furthermore, the state-of-the-art approach combining density-functional theory (DFT) with {\it ab initio} atomistic thermodynamics~\cite{Northrup1997,Reuter2001,Walle2002,Reuter2003} allows us to predict the environment-dependent (e.g., temperature and pressure dependent) binding behavior of H both on solid surfaces and in the bulk of a material employing (periodically repeated) supercells for impurities (e.g., H) contained within a finite volume of the host material. Increasing computer power and continuous improvement of the methodology facilitate a high accuracy of the predictions. 

Here, we investigate the origin of the APT-measured H residuals by combining DFT calculations on a selection of metals (Na, K, Pd, and Pt) and APT experiments on pure Na, a metal with a low-evaporation field, and pure Pt, a metal with a relatively high-evaporation field. Across several datasets, the Na APT measurements exhibit no H-related peaks, in contrast to Pt. Thermodynamic analysis based on DFT calculations allows us to determine the temperature- and pressure-dependent stability of metal surfaces in contact with H gas at the relevant vacuum conditions. Our study sheds light on the origins of H residuals in APT measurements, i.e., the detected H mainly originate from H located at the metal surface, either from contamination during specimen preparation and transfer, or during the APT measurement from adsorption of residual H$_2$ from the chamber onto the surface and migration towards the specimen's apex, making this highly dependent on the analysis conditions. These insights are critical to further optimize experimental workflows enabling the quantification of hydrogen in materials by APT. 

\section{\label{Sec: Methodology} Methodology}

\subsection{\label{m_subsec1} APT specimen preparation from Na}
Performing APT analysis requires a needle-shaped specimen in order to generate the intense electrostatic field necessary to initiate the field evaporation of the surface atoms. There are challenges inherent to the sample preparation of alkali metals (e.g., Li, Na, and K), in comparison to transition metals (e.g., Pt). Alkali metals are reactive when in contact with moisture and air (i.e., oxygen), leading to severe oxidation during the sample transfer to form NaO, which is unstable and soon reacts with H to form NaOH. These issues have so far hindered the characterization of alkali metals by APT. Here, we used a specific setup to prepare and transfer specimens that is described in detail in Ref.~\onlinecite{10.1371/journal.pone.0209211}. 

\begin{figure}[t]
\center
\includegraphics[width=0.7\columnwidth]{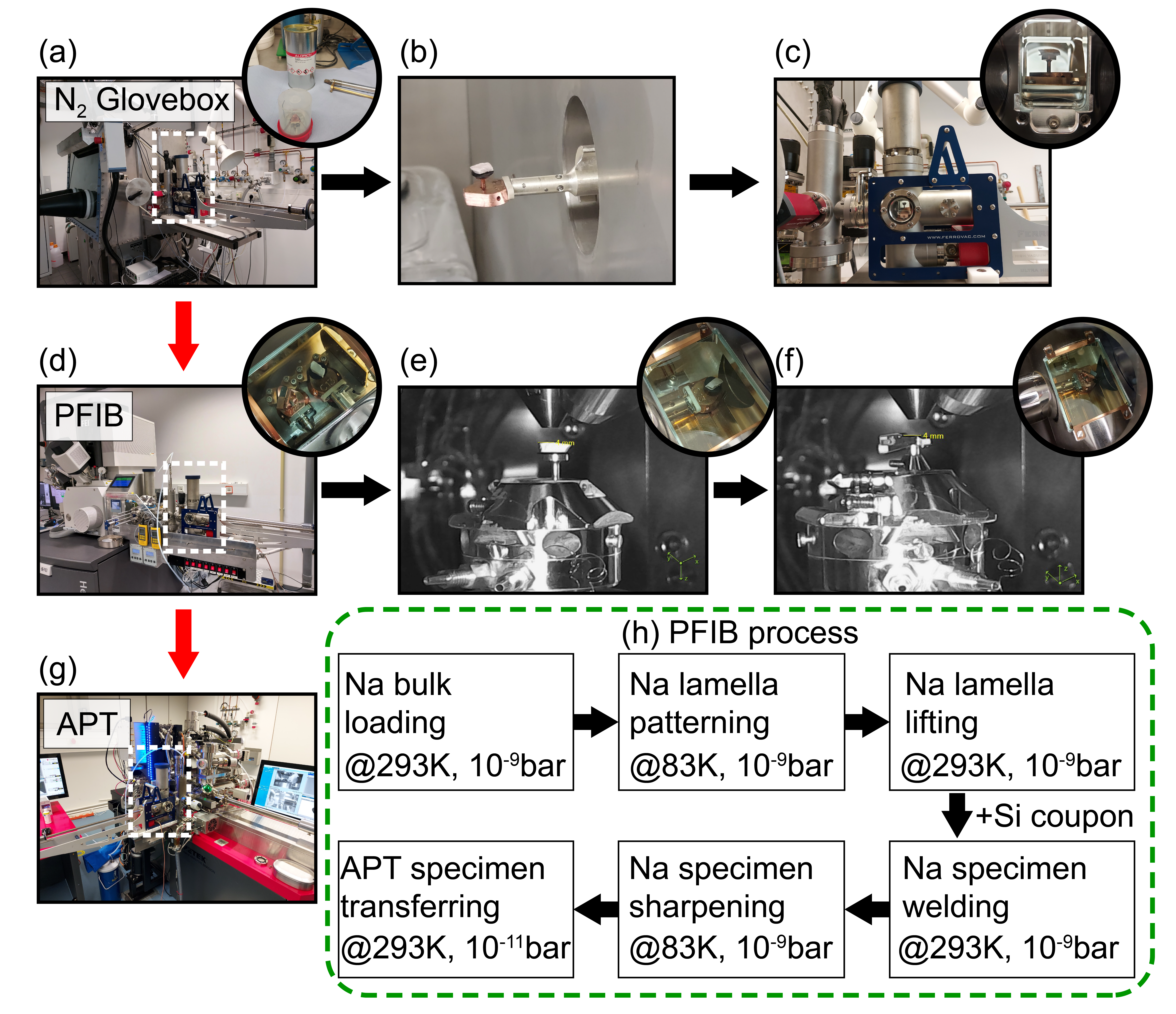}
\caption{An illustration of environmentally sensitive sample preparation for atom probe measurement. (a) The N$_2$ glovebox with the UHV suitcase (white dotted box). Inset shows the Cu-stub puck and Na bulk. (b) The sliced Na is mounted on the puck. (c) The Na in the UHV suitcase. (d) The suitcase is detached and subsequently attached to the Xe-plasma FIB. Inset shows the inside of the docking chamber. (e) The Na and (f) the Si coupon in the FIB chamber. (g) After the fabrication of the APT specimens, the suitcase is attached to APT. (h) The overall process for transferring the Na sample for APT measurement.}
\label{fig1}
\end{figure}

First, a Na sample ($>99$\,\%, Sigma Aldrich) submerged in kerosene oil was prepared inside an N$_2$-filled glovebox (Sylatech GmbH, Walzbachtal, Germany) to avoid oxidation [Fig.~\ref{fig1}(a)]. The Na sample was first sliced into a small piece ($0.5\times0.75\times0.3$\,cm$^3$) and the piece of bulk Na was attached to a flat Cu stub with adhesive carbon tape. The stub was placed in a CAMECA cryogenic APT puck (CAMECA Instruments, Madison, USA) [Fig.~\ref{fig1}(b)]. This assembly was quickly loaded into an ultrahigh vacuum (UHV) suitcase (VSN-40, Ferrovac GmbH, Zurich, Switzerland) [Fig.~\ref{fig1}(c)]. Once the pressure inside the suitcase reached $<10^{-11}$\,bar, it was detached and transported to a xenon plasma focused ion beam (FIB) /scanning electron microscope (SEM) (Helios PFIB, Thermo-Fisher, Eindhoven, Netherlands) equipped with a Ferrovac docking station [Fig.~\ref{fig1}(d)]. 

Alkali metals, and particularly Na, were reported to react strongly with the Ga used in conventional FIBs~\cite{Zachman2018}, and the heating and radiation damage caused by the ion-beam milling can lead to melting of the sample~\cite{Rubanov2001}. This explains our choice of a Xe-plasma FIB to limit the ion-beam damage and the reactivity of the implanted ions, as Xe is more inert compared to Ga~\cite{JMayer2007,BURNETT2016119}. Moreover, the cryo-stage is implemented to avoid uncontrolable melting of the Na sample during the FIB process (see Fig.~S1 in the Supporting Information).

Figure~\ref{fig1}(e) shows the Na sample inside the FIB chamber. The cryo-stage (Gatan C1001, Gatan Inc., Pleasanton, CA, USA) was pre-cooled to $-189$\,$^\circ$C by cold gaseous nitrogen. A wedge-shaped lamella from the Na bulk was prepared using the lift-out protocol described in Ref.~\onlinecite{THOMPSON2007131}. Clean trenches were milled on the Na surface for the lift-out process. In order to prevent the condensation of gaseous platinum deposition precursor molecules from the lift-out process (e.g. methylcyclopentadienyl trimethyl platinum), the stage was warmed up to room temperature to weld the wedge onto the micromanipulator.

\begin{figure}[t]
\center
\includegraphics[width=0.9\textwidth]{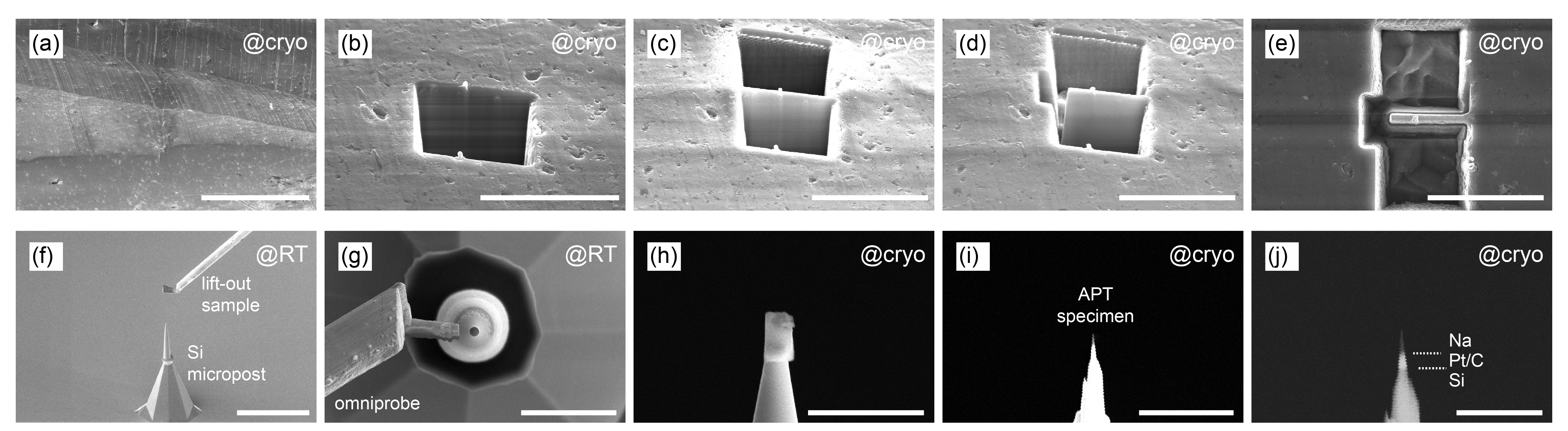}
\caption{APT specimen fabrication process: (a) the as-received Na bulk sample imaged at cryo-temperature. Trenches were milled on (b) the front and (c) the back sides of the interesting region with a width of $<2$\,$\mu$m in 52\,$^\circ$ perpendicular to the ion beam column. (d,e) The L-shape horizontal cut was made at the bottom and the left side of the interesting region in 0\,$^\circ$. (f,g) The stage temperature was heated up to room temperature and the sample was welded with a FIB-Pt deposition on a Si micro-tip. (h) After the welding, the stage was cooled down again to cryo-temperature, and (i) annular milling from the top was performed until the apex radius was below 100\,nm with no pore. (j) BSE image of the final APT specimen. For the Xe cleaning process, the final beam voltage of 5\,kV and 10\,pA were used to etch remaining residuals on the specimens. scale bars: (a) 100\,$\mu$m, (b)-(e) 20\,$\mu$m, (f) 100\,$\mu$m, (g)-(i) 5\,$\mu$m, (j) 2\,$\mu$m}
\label{fig2}
\end{figure}

Scanning electron micrographs, taken at 5\,kV and 1.6\,nA, for each of the successive steps of the preparation are shown in Fig.~\ref{fig2}(a–j). A low-electron-dose image was taken since the alkali material is also sensitive under electron beam (e.g., Li battery reacts with the electron beam~\cite{Yuan2017}). After the milling process, no open pores or cracks were observed in the SEM and the back-scattered electron (BSE) image in Fig.~\ref{fig2}(j) shows significant contrast difference among Na, Pt deposited during welding, and the Si micro-pillar regions.

After the lift-out was performed, the APT puck with the Cu stub was moved back to an intermediate UHV storage chamber part of the Ferrovac docking station. Another APT puck with a commercial Si coupon was inserted in the FIB (previously placed in the intermediate UHV storage chamber) [see Fig.~\ref{fig1}(f)]. The Na lamella was welded to several Si supports using standard Pt precursor (Methyl cyclopentadienyl trimethyl platinum). The cryo-stage was then cooled again to $-189$\,$^\circ$C, and the stage was tilted to 52\,$^\circ$ to be perpendicular to the ion beam column. Progressively, smaller annular milling patterns were used to sharpen the Na into specimens suitable for APT (e.g., tip-diameter less than 100\,nm) [see Fig.~\ref{fig2}(i)]. After the final milling, the cryo-prepared specimens were transferred from the PFIB chamber to the UHV suitcase and subsequently transferred into the CAMECA LEAP (local electrode atom probe) 5000 HR [Fig.~\ref{fig1}(g)]. To summarize, the overall process is shown in Fig.~\ref{fig1}(h). For the Pt specimen, the same protocol was conducted as for Na. The only difference was that Pt bulk was loaded through the FIB intermediate chamber, not the N$_2$ glovebox.

\subsection{\label{m_subsec2} APT measurement}
Atom probe data were acquired in laser-pulsing mode with a pulse energy of 70\,pJ and rate of 50\,kHz at 1\,\% evaporation rate by adjusting an applied DC voltage. The base temperature was set throughout the measurement to 30\,K and 90\,K, respectively. The Na and Pt specimens that were field evaporated at 90\,K base temperature are labelled as Na$_{\rm 90K}$ and Pt$_{\rm 90K}$ whereas the Na specimen at 30\,K is labeled as Na$_{\rm 30K}$. The chamber pressure was in the 10$^{-14}$\,bar. The 3D data reconstruction, data analysis, and visualization were performed using AP SUITE software version 6.1.

\subsection{\label{m_subsec3} Computational details}
All DFT calculations are performed using the Vienna Ab initio Simulations Package (VASP)~\cite{Kresse1996,Kresse1996a} with the projector augmented wave (PAW) approach~\cite{Bloechl1994}. The kinetic-energy cutoff employed for the plane-wave basis set is 500\,eV. The generalized gradient approximation (GGA) is used for the exchange-correlation approximation~\cite{Perdew1996,Perdew1997,Hammer1999} (see Table~S1 in the Supporting Information for details of the used approximations). Electronic and ionic relaxations are carried out until the total energy convergence is less than 10$^{-5}$\,eV, respectively, 10$^{-4}$\,eV. 

A repeated slab approach is used to study the most favorable surface planes for body-centered cubic (BCC) and face-centered cubic (FCC) metals, specifically Na(110), K(110), Pd(111) and Pt(111). The periodically repeated slabs are decoupled by adding a vacuum region and applying the dipole correction scheme~\cite{Neugebauer1992}. A detailed description of the computational setup for each metal surface calculation (e.g., vacuum thickness, supercell size, $k$-point mesh, and slab thickness) is given in Table~S1 in the Supporting Information. We employ surface cells larger than the $p({1\times1})$ cell to account for various coverages ($\Theta$) up to 1\,ML (monolayer) (e.g., from $\Theta=1/12$\,ML to 1\,ML for H on Pt(111)). Defining the coverage as the ratio between the number of adsorbate atoms and the number of metal atoms in the top surface layer, 1\,ML is reached for an equal number of adsorbate and surface metal atoms. 

\subsection{\label{m_subsec4} Surface thermodynamics}

The binding energy of hydrogen on the metal surface, $E_{\rm b}$, with respect to a H$_2$ molecule is calculated as

\begin{equation}
E_{\rm b} = (E^{\rm H-surf}_{\rm tot} - E^{\rm clean-surf}_{\rm tot} - \frac{1}{2}\cdot N_{\rm H}\cdot E^{\rm H_2}_{\rm tot})/N_{\rm H} \quad ,
\label{eq1}
\end{equation}
where $E^{\rm H-surf}_{\rm tot}$, $E^{\rm clean-surf}_{\rm tot}$, and $E^{\rm H_2}_{\rm tot}$ are the DFT calculated total energies of a metal surface with and without (i.e., of a clean metal surface) adsorbed H and the H$_2$ molecule, respectively. $N_{\rm H}$ is the number of hydrogen atoms adsorbed on the surface.

To account for the stability of the metal surface in a H atmosphere as a function of the temperature and pressure, the change in the Gibbs free energy of each surface phase with respect to a H-free metal surface is calculated as,

\begin{equation}
\Delta G^{\alpha} = [E^{\rm H-surf}_{\rm tot} - E^{\rm clean-surf}_{\rm tot} - N_{\rm H}\cdot\mu_{\rm H}(T, p) - T\cdot S_{\rm conf}]/A \quad ,
\label{eq2}
\end{equation}
where $\mu_{\rm H}(T, p)$ is the chemical potential of hydrogen, which is a function of temperature ($T$) and pressure ($p$), $A$ is the area of the surface cell and $S_{\rm conf}$ is the configurational entropy of the surface atoms. The later is approximated by $-k_{\rm B}\cdot [\Theta\cdot\ln\Theta + (1-\Theta)\cdot\ln(1-\Theta)]$, where $k_{\rm B}$ and $\Theta$ are the Boltzmann constant and the coverage of the adsorbates, respectively.

The H chemical potential [$\mu_{\rm H}(T, p)$] is evaluated as follows~\cite{Northrup1997,Reuter2001,Rogal2007},
\begin{equation}
\label{eq3}
\begin{split}
\mu_{\rm H}(T, p) & = \frac{1}{2}E^{\rm H_2}_{\rm tot} + \frac{1}{2}E^{\rm H_2}_{\rm ZPE} + \Delta \mu_{\rm H}(T, p)  \\ 
\Delta \mu_{\rm H}(T, p) & = \frac{1}{2}[H_{\rm H_2}(T, p^0) - H_{\rm H_2}(0 {\rm K}, p^0)] 
- \frac{1}{2}T[S_{\rm H_2}(T, p^0) - S_{\rm H_2}(0 {\rm K}, p^0)] \\
&  + k_{\rm B}T\ln{\frac{p}{p_0}} \quad ,\\
\end{split}
\end{equation}
where $\frac{1}{2}E^{\rm H_2}_{\rm tot}$ and $E^{\rm H_2}_{\rm ZPE}$ are the total energy at $T=0$\,K and the zero-point energy of a hydrogen molecule. Our calculated value for $E^{\rm H_2}_{\rm ZPE}$ is 0.273\,eV. $\Delta \mu_{\rm H}(T, p)$ contains the temperature- and pressure-dependent free energy contributions. Assuming that H$_2$ gas behaves like an ideal gas, the temperature dependence at standard pressure (i.e., $p^0 = 1$\,atm) is evaluated using tabulated values for enthalpy ($H$) and entropy ($S$) at finite temperature~\cite{JANAF} and the relationship, $G = H - TS$. 

\section{\label{Sec: Results} Results}

\subsection{\label{r_subsec1} Atom probe results}
\begin{figure}[t]
\center
\includegraphics[width=0.6\columnwidth]{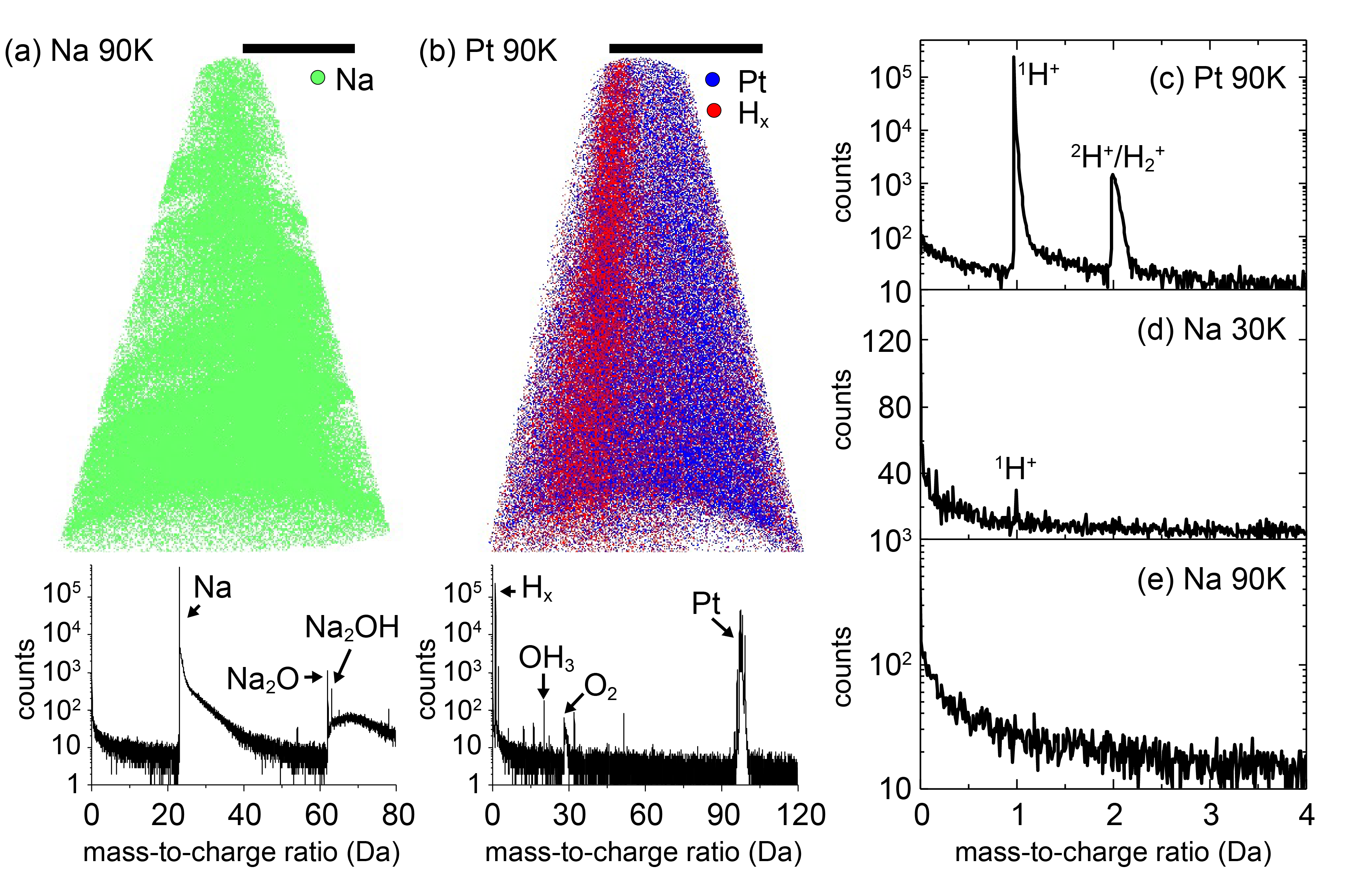}
\caption{3D atom maps and corresponding mass spectra of (a) Na$_{\rm 90K}$ and (b) Pt$_{\rm 90K}$ samples [scale bars are (a) 50 and (b) 20\,nm]. For the mass spectra analyses, 2.5\,M ions were extracted from each atom map. Local region in the mass spectra of (c) Pt$_{\rm 90K}$, (d) Na$_{\rm 30K}$, (e) Na$_{\rm 90K}$.}
\label{fig3}
\end{figure}

Figure~\ref{fig3}(a) and (b) show the 3D atom maps and corresponding mass spectra acquired from the Na$_{\rm 90K}$ and the Pt$_{\rm 90K}$, respectively. All samples were in the laser pulsing mode (background levels $<10$\,ppm/nsec). While the Pt measurement was smooth, there were several micro-fractures during the Na measurement. We tried high voltage pulsing for both materials, however the level of background signals was higher than would be acceptable ($>1000$\,ppm/nsec) and the Na specimens failed after the collection of only less than 52,000 ions at a pulsed voltage of 10-percentage. The sizes of acquired dataset for Na$_{\rm 90K}$ and Na$_{\rm 30K}$ APT measurements were $>10$\,M ions. In the acquired mass spectrum of the Na$_{\rm 90K}$, strong peaks appear at 23, 62, and 63\,Da correspond to Na, Na$_2$O, and Na$_2$OH, respectively. These peaks associated to residual -O and -OH  are frequently observed in experiments following cryo-UHV transfer and can be also associated to low level of frosting on the specimens following preparation ~\cite{khanchandani2021laserequipped,10.1371/journal.pone.0209211}. The composition of the whole Na$_{\rm 90K}$ sample is 99.006\,\% Na, 0.987\,\% O, and 0.007\,\% H following peak decomposition \cite{London2019}. Figure~\ref{fig3}(c), (d) and (e) display the section of the mass spectrum for the hydrogen peaks of the analyses of Pt$_{\rm 90K}$, Na$_{\rm 30K}$, and Na$_{\rm 90K}$, each dataset containing ${2.5\times10^6}$ identified ions. 

No peak pertaining to H species, at 1, 2 or 3\,Da, is visible above the level of background in the mass spectrum from the analysis of the Na$_{\rm 90K}$ plotted in  Figure~\ref{fig3}(e), which corresponds to the lowest electric field conditions across the data reported herein. For the Na$_{\rm 30K}$ analysis, a small peak at 1\,Da is visible, whereas for Pt, strong peaks at 1 and 2\,Da are clearly resolved. 

\subsection{\label{r_subsec2} Surface thermodynamics and electronic structures}

\subsubsection{H Binding on metal surfaces}
\begin{figure}[t]
\center
\includegraphics[width=0.8\columnwidth]{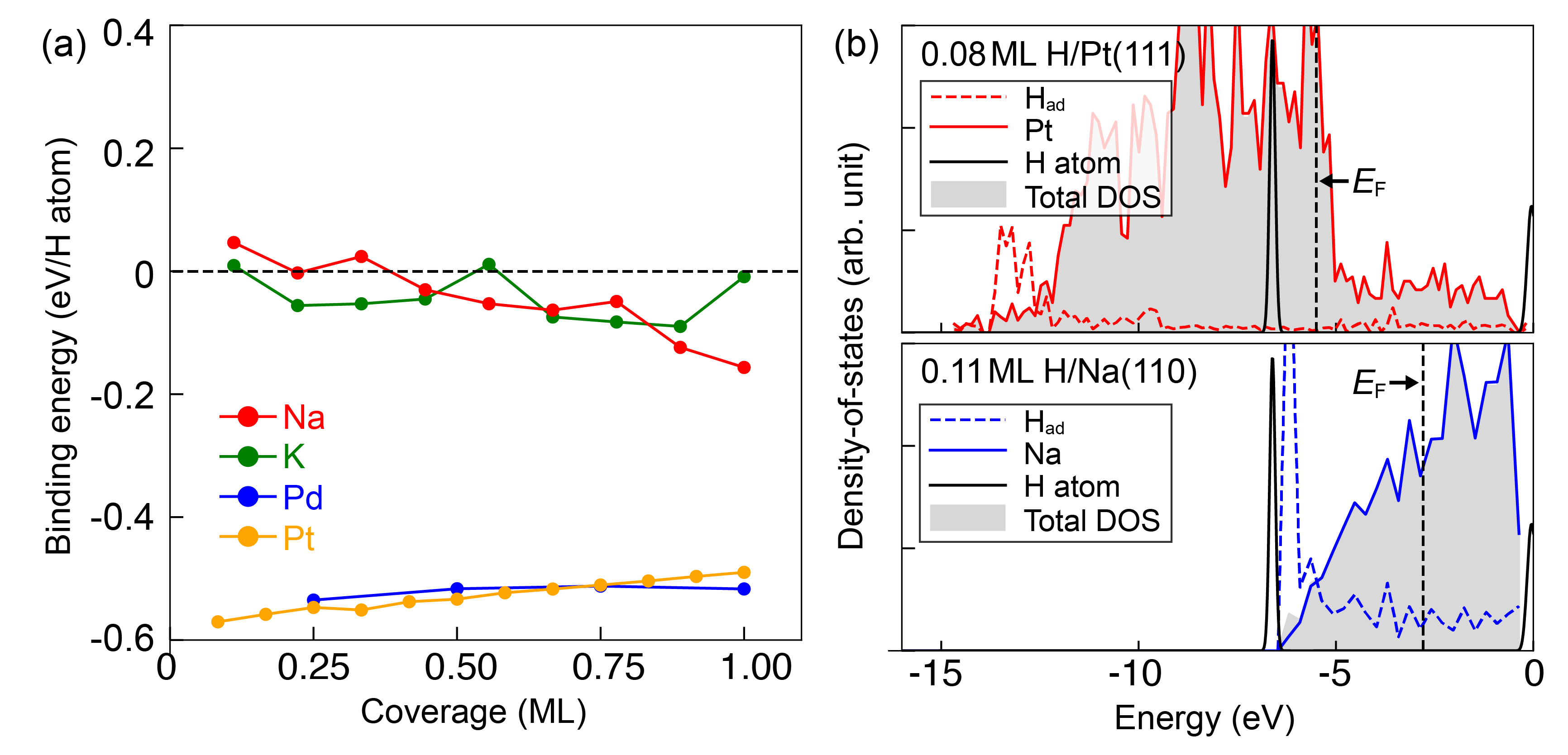}
\caption{
(a) Binding energies, $E_{\rm b}$, of H adsorbates on the four metal surfaces for several adsorbate coverages up to 1\,ML. Each colored solid line corresponds to a different metal (i.e., red - Na, green - K, blue - Pd, and orange - Pt). 
(b) The density-of-states (DOS) of (upper panel) 0.08\,ML H-Pt(111) and (bottom panel) 0.11\,ML H-Na(110) surfaces, where the total, Pt atom-resolved, and H adatom-resolved DOS are indicated by a grey region, a solid colored line, and a dashed colored line, respectively. The DOS of a single H atom in a box is shown as a black solid line. The energy is aligned to the vacuum level set to zero. The position of the Fermi level $E_{\rm F}$ is indicated by a vertical black dashed line.
}
\label{fig4}
\end{figure}

To elucidate the surface stability of metals in a H-containing atmosphere, we first identify the favorable chemisorption sites for H on the respective metal surface. Based on literature reports for FCC metals [e.g., Pd(111)~\cite{Paul1996,Lovvik1998} and Pt(111)~\cite{Yan2018}] we consider in the following the FCC-hollow sites (i.e., a triply coordinated binding site, on which a next layer metal atom in the continuation of an FCC-stacking sequence would be found) for Pd and Pt. For Na(110) and K(110) we tested various binding sites, specifically top, long-bridge, short-bridge, and hollow sites, because of a lack of previous studies. Increasing the surface coverage from of H 0.11\,ML to 1\,ML, we consistently find that H prefers to bind to the hollow sites. 
Having identified the most favourable adsorption sites, we calculate the H binding energy ($E_{\rm b}$) on each of the metal surfaces using Eq.~\ref{eq1}. The result is shown in Fig.~\ref{fig4}(a) for all considered coverages. The H-binding energies $E_{\rm b}$ on the surface of the FCC metals (Pd and Pt) are roughly half eV more negative (i.e., more stable) than those on the Na and K surfaces. This clearely indicates a much stronger binding on Pd(111) and Pt(111), which we find to be in the order of $-0.6$ to $-0.5$\,eV/atom over the whole range of considered H coverages up to 1\,ML, in agreement with previous theoretical studies~\cite{Roudgar2003,HANH2014104}. In contrast, the binding energies on the Na(110) and K(110) surfaces are substantially weaker and in a range of $0.1$ to $-0.2$\,eV. The positive binding energy calculated for the lowest coverage ($\Theta = 0.11$\,ML) on the Na(110) surface indicates that H adsorption is thermodynamically unstable, as forming a H$_2$ molecule is an exothermic reaction. With increasing coverage the H binding energy decreases. This is related to the propensity for alkali hydride formation (i.e., NaH and KH) at the surface, as argued in a previous study~\cite{HJELMBERG1979539} regarding the phase transition from bulk Na to NaH upon exposure to large amounts of H$_2$ gas. In fact, we actually observe the formation of surface hydride [i.e., rocksalt NaH(100)] on the Na(110) surface at 1\,ML coverage in our DFT calculation.

There has been a clear consensus about the high solubility of H in Pt and Pd bulk since the early 1930's~\cite{Lacher1937}, which implies that diffusion of H atoms from the surface into the bulk region is likely. It suggests that H is simultaneously present both at the surface and in the bulk. The lack of studies for alakali metals regarding this point prompted us to calculate the binding energy of a H interstitial defect in the sub-surface region of Na. We considered different coverages for a H in a tetrahedral site of the 1st, 2nd up to the 5th sub-surface layer of the Na(110) surface. As shown in Fig.~S2 in the Supporting Information, all the binding energies at sub-surface sites are larger than the surface binding energies at the corresponding coverage, which indicates an endothermic binding reaction compared to the surface binding. Furthermore, the further an interstitial H atom is away from the surface, the less favorable its binding energy becomes. In fact, binding energies in deeper layers are close to the formation energy of a H bulk interstitial defect (in a tetrahedral site) as calculated using a Na $p({4\times4\times4})$ bulk supercell [i.e., $E_{\rm f}({\rm H}) = 0.13$\,eV/atom]. This trend consistently demonstrates that H migration from the surface to the sub-surface region and subsequently to the bulk is unlikely in the Na metal, in contrast to the Pd and Pt systems.

To better understand why H binds less strongly on alkali surfaces compared to transition metal surfaces, we analyse the electronic structure, selecting the 0.08\,ML H-Pt(111) and 0.11\,ML H-Na(110) surfaces as representative cases. Their corresponding binding energies are 0.05 and $-0.57$\,eV/H atom, respectively, and their density-of-states (DOS) are shown in the upper and the bottom panels of Fig.~\ref{fig4}(b). The total DOS (grey region), the atom-resolved DOS (solid colored line for metal atoms and dashed colored line for the H adatom) and the DOS of a single H atom in a box (black solid line) are aligned with respect to the vacuum level and set to zero. For Pt (the upper panel) we observe that the localized H 1$s$ states in vacuum (black solid line) seen near $-7$\,eV largely overlap with the DOS of the Pt atoms (red solid line). This leads to a  strong hybridization between the H $s$ and Pt $d$ states resulting in H $s$ bonding states at lower energy (see a peak in the red dashed line near $-14$\,eV). Such a strong $s$-$d$ hybridization is also confirmed by the analysis of the electron density differences shown in Fig.~S3(c) and (d) in the Supporting Information. A strong bond between H and Pt surface atoms is in agreement with the conventional model explaining bonding between H and transition metal surfaces, the so-called $d$-band model~\cite{Norskov2011}. 

In the case of the Na system [the bottom panel of Fig.~\ref{fig4}(b)], the $s$ states of the isolated H atom in vacuum are located at a tail of the Na $s$ states, leading to only a relatively small overlap. This hinders a $s$-$s$ hybridization, i.e., there is hardly any change in the DOS of H$_{\rm ad}$ (blue dashed line). Instead, a charge transfer from the Na surface atoms towards the H atom occurs, resulting in a negatively charged H adsorbate atom, as confirmed by the electron density difference [Fig.~S3(a) and (b) in the Supporting Information]. The clear difference in the bonding nature of H on Na and Pt surfaces explains the difference in H binding energies.

\subsubsection{Phase diagrams for H-metal surfaces}

Having understood the binding behavior of H on these metal surfaces, we can now utilize our calculations to construct phase diagrams and account for the impact of temperature and pressure on the surfaces' stability. For this we need to evaluate the Gibbs free energy change, $\Delta G$, of H-covered metal surfaces with respect to the clean surface calculated  as a function of the chemical potential of H based on Eq.~\ref{eq2}. As shown in Fig.~S4 in the Supporting Information, $\Delta G$ of H-covered surfaces becomes lower than that of the clean surface for all metals as the H chemical potential increases. This indicates that H-covered surfaces with higher H coverage become energetically more favorable at higher H chemical potentials. Based on Eq.~\ref{eq3} we can explicitly evaluate the H chemical potential for any given set of temperature and pressure, which allows us to construct phase diagrams showing the thermodynamically most favorable surface phases at the given conditions, which are depicted as color regions in Fig.~S5 in the Supporting Information. To simplify the phase diagrams, only phase boundaries between the H-free clean surface and the first H-covered surface that becomes stable are shown in Fig.~\ref{fig5}(a) for each metal. The regions to the right and left of each solid line correspond to the stable region for the H-free and the H-covered surface, respectively. Similar to the binding energies, the four solid lines can be split in two groups (i.e., alkali metals and FCC transition metals). Based on this phase diagram, we expect that the Pt and Pd surfaces will be easily contaminated by H adsorbates even at moderate conditions (e.g., $T = 300$\,K and $p = 1$\,bar). The clean surface becomes favorable only at extreme conditions [e.g., $p < 10^{-9}$\,bar at 300\,K for Pt(111)]. The situation is different for the alkali metals, where a comparatively higher resistances to surface contamination with H is observed. For example, at $p = 10^{-1}$\,bar and $T = 300$\,K H-free clean surfaces are thermodynamically more favorable than H-covered surfaces. To connect the theoretical phase diagrams to the experiments, the experimental conditions are shown: colored crosses specify conditions at which a metal tip is prepared (brown), the prepared tips are transferred to the APT equipment (cyan), and the APT measurement is performed in a vacuum chamber (pink). The effect of the electric field present during the operation of APT analysis is also taken into account: the dashed colored lines are the phase boundaries for the Na and Pt surfaces which shift in the presence of the electric field and resulting changes in the dipole moments of the surface phases. Based on the constructed phase diagrams we rationalize the detection of the H atoms in the APT-measured mass spectra in the following section.  

\section{\label{Sec: Discussions} Discussions}

\subsection{Origin of the detected hydrogen}
Our APT analyses of Na at 30 and 90\,K show no measurable amount of H, potentially contradicting the common view regarding the origin of the background H from ionization of residual gases. This contrasts with the results for Pt for which substantial amounts of H and H$_2$ are detected. In this section we discuss possible scenarios for H contamination of metals and reflect on the a long-standing debate on the origin of H species detected in APT-measured mass spectra [e.g., Fig.~\ref{fig3}(c)]. 

It is commonly accepted that residual H$_2$ molecules are still present in a vacuum chamber even at extremely low pressure and temperature. This is in part due to the relatively high content of hydrogen within the stainless steel most vacuum chamber are made of. Residual H$_2$ molecules can then ionize and dissociate due to the intense electric field and/or laser pulse during APT operation. This leads to the detection of H species without any direct interaction with the specimen, simply associated to typical field ionization as encountered in field-ion microscopy for instance. The ionization potential of H$_2$ is 15.4\,eV, and it has been used as an imaging gas, in particular for silicon~\cite{Melmed1975,Koelling2013a}. 

H species present inside the specimen are detected during APT measurements. Even if the presence of H atoms in a sample can be undesired, contamination may occur during the specimen preparation and transport, or during the measurement itself. The importance of controlling the temperature during specimen preparation was recently pointed out for several alloy systems~\cite{Chang2019, Lilensten2020}, and in particular for materials systems that are known hydride-formers~\cite{Mouton2021}. Breen et al. demonstrated the strong ingress of hydrogen arising from the specimen preparation by electrochemical polishing~\cite{BREEN2020108}. 

\begin{figure}[t]
\center
\includegraphics[width=0.9\columnwidth]{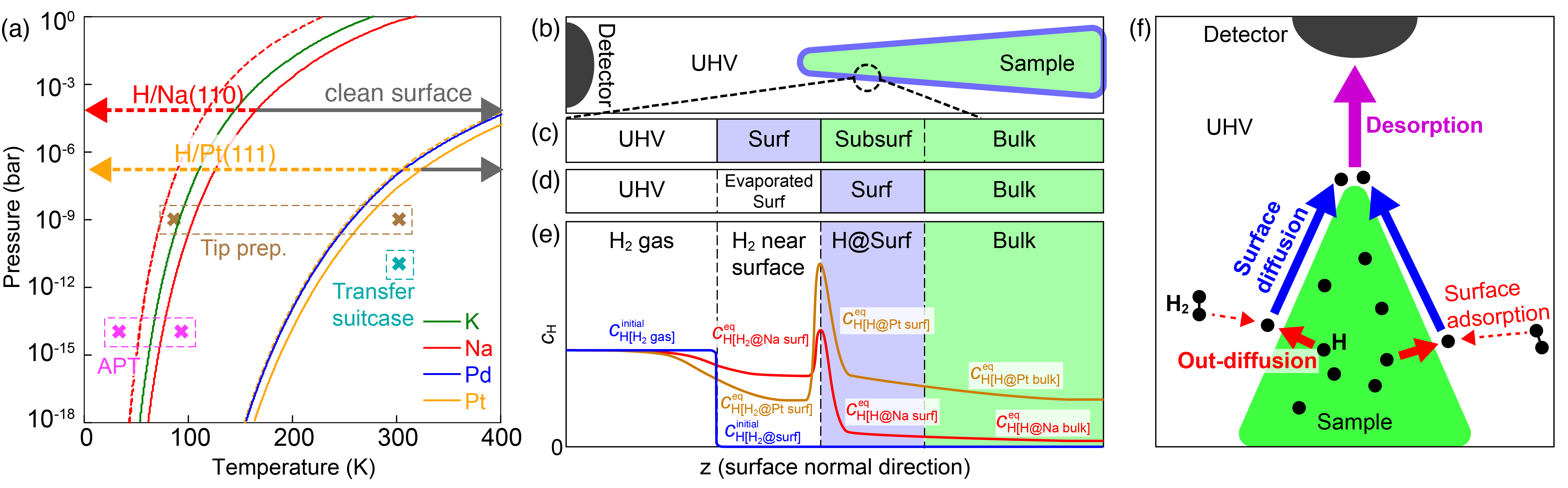}
\caption{
(a) Surface phase diagram for metal surfaces in equilibrium with a surrounding H$_2$ gas. Each colored solid line indicates the phase boundary between the clean surface (a region to the right of the line indicated by color-dashed arrows) and the H-covered surface (region to the left, as indicated by grey arrows) for each metal. Corresponding phase boundaries for the Na and Pt surfaces, shifted as a consequence of the effect of the electric field and modified dipole moment of the surface phases (see Section.~\ref{Sec: Discussions} for details), are shown as coloured dashed lines. Pink, brown, and cyan crosses show the experimental conditions for APT measurements, tip preparation, and tip transportation, respectively. 
The schematic picture (b) illustrates the environment of the APT equipment and zoomed-in surface regions of a tip in a contact with H$_2$ gas (c) before and (d) after surface metal layers evaporation during an APT measurement, and (e) the H$_2$ molecules and H atoms concentration profiles corresponding to each geometric region in (d). A blue solid line indicate the initial H concentration. Red and dark orange lines illustrate the H concentration at the thermodynamic equilibrium for Na and Pt, respectively.
(f) This schematic illustrates the origin of the H residuals in the APT measurement. Black-filled circles and a green shape represent H atoms and the metal specimen, respectively. Each arrow conceptually shows corresponding diffusion paths for H diffusing from the bulk (red), H surface-diffusing towards the specimen's apex (blue), and desorbed H from the apex (purple). A relatively very slow diffusion of H$_2$ molecules in the UHV is illustrated by dashed red arrows.
}
\label{fig5}
\end{figure}

Here, we prepared specimens either at 90\,K or at 300\,K, but at a fixed pressure of approx. 10$^{-9}$\,bar inside the focused ion beam. The specimens were then transported into the atom probe by using an ultra-high vacuum transfer suitcase, being maintained at approx. 300\,K and 10$^{-11}$\,bar. According to our phase diagram [Fig.~\ref{fig5}(a)], both the Na and Pt surfaces will not become covered by H during this process, because the respective H-free clean surfaces are thermodynamically favored. However, Pt samples can become contaminated by H atoms when the specimen is created in the vacuum chamber, especially at the conditions of 90\,K and 10$^{-9}$\,bar, while this is unlikely for a Na surface under similar conditions. Furthermore, given the high affinity of H and Pt discussed in the previous section, we cannot neglect the possibility of H diffusion from a H contaminated surface of a Pt sample into either its sub-surface or bulk region. This mechanism was pointed out to be responsible for the large ingress of H in titanium specimens during specimen preparation~\cite{Chang2019} under similar preparation conditions. Therefore, there is a high likelihood that the Pt specimen has already absorbed H atoms even before the APT measurement commences, which can be excluded in the case of Na.

An alternative scenario in which surfaces of samples are contaminated during the actual APT measurement is also conceivable, assuming that the metallic specimen is initially devoid of H. The detected H-related signal can originate from the binding of H to the specimen's surface during the APT measurement as a consequence of interactions between the H$_2$ gas and the surface metal atoms. First we note that from a thermodynamic perspective H-covered Pt surfaces are more stable than the clean surface at the conditions of our APT experiments (i.e., $10^{-14}$\,bar and 30$\backsim$90\,K). This is shown in Fig.~\ref{fig5}(a) and suggests a high possibility of contamination. However, the Na phase boundary [red line in Fig.~\ref{fig5}(a)] intersects the region of specified APT conditions. This means that the clean Na surface is stable against H chemisorption even at the relatively higher temperatures, e.g., 90\,K, whereas H-covered surfaces are expected to form at the lower temperature, e.g., 30\,K. This is supported qualitatively by the APT measurements reported in Fig.~\ref{fig3}.

\subsection{Surface contamination during analysis}
During an APT measurement, the surface atoms are progressively field evaporated from the specimen's surface, fly through the ultra-high vacuum and are collected by the particle detector, as illustrated in Fig.~\ref{fig5}(b). Therefore, it is necessary to also consider kinetic scenarios, which can be involved in surface contamination by H. The field evaporation of the specimen's outermost surface layer, as illustrated in Fig.~\ref{fig5}(c), exposes a new surface layer that was previously in the sub-surface region, as depicted in Fig.~\ref{fig5}(d). This means that the space where the original surface layer was found has now become empty. Therefore, there is not only a lack of metal atoms in this region, but it is also void of H$_2$ molecules. The later (i.e., H$_2$) have to diffuse from the vacuum through this space towards the surface. The fundamental force driving this diffusion (i.e., of H$_2$ from the vacuum towards the surface) is the gradient of the H concentrations (i.e., $c_{\rm H}$), as it strives to achieve thermodynamic equilibrium. Consequently, the diffusive flux ($J$) of H$_2$ can be modeled based on a linear law as $J = - D\cdot\nabla c_{\rm H}$, where $D$ is the diffusion coefficient and $\nabla c_{\rm H}$ is the H concentration gradient. 

This process is schematically illustrated in Fig.~\ref{fig5}(e). The corresponding H concentration, denoted as $c_{\rm H [{\alpha}]}$, where $\alpha$ denotes the source for H, which in the presently discussed case depends on the geometric positions, e.g., H$_2$ in vacuum, H$_2$ in the vicinity of the surface, H adsorbed at the surface, or H in the bulk of the material. The initial H concentration is indicated by the blue solid line in Fig.~\ref{fig5}(e) in which H species are present only in the region far away from the surface again assuming that the metallic specimen does not initially contain any H atom. The processes related to the formation of H-contaminated surfaces can be decomposed into four steps
\begin{itemize}
\item (i) H$_2$ molecules diffuse from the vacuum to the region near the surface due to the concentration gradient between $c^{\rm initial}_{\rm H[H_{2}\,gas]}$ and $c^{\rm initial}_{\rm H[H_{2}@surf]}$; 
\item (ii) the diffused H$_2$ molecules near the surface dissociate into two H atoms by overcoming an energy barrier; 
\item (iii) a H-contaminated surface forms, if H$_2$ molecules dissociate and resulting H atoms are chemisorbed on the metal surface; 
\item (iv) H atoms migrate from the surface to the bulk region if the H surface chemical potential at the given surface concentration is above the bulk chemical potential at the given bulk concentration.
\end{itemize}

All steps from (i) to (iv) are material-dependent processes, because all concentrations, except for H$_2$ gas in the vacuum far away from the surface (i.e., $c_{\rm H[H_{2}\,gas]}$), are related to the material itself. For example, the concentrations at thermodynamic equilibrium conditions are qualitatively depicted in Fig.~\ref{fig5}(e) for Na with a solid red line and for Pt with a solid dark orange line. Once the thermodynamic equilibrium is achieved, $c^{\rm eq}_{\rm H[H@Pt\,surf]}$ would be substantially larger than $c^{\rm eq}_{\rm H[H@Na\,surf]}$, since the stability of H-contaminated surfaces is higher for Pt than for Na. This implies a much more active exchange of H species for Pt. 

Furthermore, H migration to the bulk region is not expected for Na surfaces but it is for Pt, so that $c_{\rm H[H@Pt\,bulk]} >> c_{\rm H[H@Na\,bulk]}$. At the conditions corresponding to phase boundaries, shown as solid colored lines in the phase diagram [Fig.~\ref{fig5}(a)], surfaces in contact with H$_2$ molecules and H-contaminated surfaces are in equilibrium and would coexist. However, as a fresh metal surface becomes exposed following field evaporation of the surface layer(s), $c_{\rm H[H@Na\,surf]}$ will be much smaller than $c_{\rm H[H_{2}@Na\,surf]}$, which precludes an exchange of H species between the surface and the environment. In other words, under the conditions of an APT measurement, denoted as pink crosses in Fig.~\ref{fig5}(a), the chances for forming a H-contaminated surface will be much higher for Pt surfaces than for Na surfaces. 

\subsection{Influence of the electric field}

Last but not least, since APT measurements are carried out in the presence of large electric fields (e.g., 1$\backsim$4\,V/{\AA})), the impact of such a large electric field on the surface stability has to be considered. 

The electric field is known to have a substantial influence on the distribution of detected hydrogen~\cite{Tsong1983,Chi-fongAi1984, Mouton2019}. The content of H, presumably originating from the residual gas from the chamber, is expected to increases as the electric field decreases~\cite{Sundell2013}. In contrast, Andren and Rolander pointed to an influence of the electric field on the hydrogen adsorption behavior and hence suggest a change in the detection of H level as a function of the base temperature~\cite{Andren1992}. For relatively low fields, hydrogen adsorbed on the surface is field evaporated alongside one of the host-metal atoms forming, as reported for Al for instance, AlH, AlH$_2$, AlH$_3$~\cite{Nishikawa1983a}. The bonding of the metal atoms with gaseous species has been reported to favour the field evaporation at lower electric fields~\cite{Muller1965}, and their detection should hence not be seen here.

The evaporation field of Na is estimated using the classical image hump model for field evaporation  to be in the range of 1.1\,V$\cdot${\AA}$^{-1}$~\cite{Tsong1978a}. No experimental value was ever reported. There are theoretical estimates of the evaporation field for Na adsorbates on Al and W that are also typically in this range of 0.6$\backsim$0.8\,V$\cdot${\AA}$^{-1}$~\cite{Kahn1976,Neugebauer1993}. Under such low fields, typically peaks appear at 1, 2, and 3\,Da in most analyses of metallic samples. These peaks are not observed here (cf. Fig.~\ref{fig3}).

Although well-established approaches exist to explicitly include the electric field in DFT surface calculations~\cite{PhysRevLett.124.176801}, they come with large computational costs. Therefore, in the following we utilize as a first approximation to account for the thermodynamic effect of the electric field only free energy contributions, $\Delta U$, due to the presence of an electric field and use the equation $\Delta U^{\alpha} = -\Delta\mu_{\rm dipole}^{\alpha}\mathcal{E}/A$. Here $\Delta \mu_{\rm dipole}^{\alpha}$ is the dipole moment difference of a surface phase $\alpha$ with respect to the H-free surface, $\mathcal{E}$ is the electric field applied during the APT measurement, in the range 0.5$\backsim$5\,V$\cdot${\AA}$^{-1}$, and $A$ is the surface area of the used slab models. By adding $\Delta U^{\alpha}$ to the Gibbs free energy changes calculated from Eq.~\ref{eq2}, we can evaluate the field dependent phase boundary positions for Na and Pt, assuming fields of 2 and 4\,V$\cdot${\AA}$^{-1}$ respectively, shown as dashed colored lines in Fig.~\ref{fig5}(a). For both Na and Pt, the stronger surface dipole (compared to the pristine surface) due to the presence of bound H atoms at the surface leads to a reduction of $\Delta G$ by up to 10\,meV/{\AA}$^2$ for the highest H coverage. This results in a destabilization of H-covered surfaces. Therefore, the phase boundaries for the Na and Pt cases, shown as red and orange dashed lines in Fig.~\ref{fig5}(a), are shifted towards the left, i.e., increasing the region in which the respective clean surface is stable. Comparison between the shifted phase boundaries and the actual conditions of the APT measurement consistently imply, from a thermodynamic standpoint, a low probability of H contamination of Na samples, even in the presence of a large electric field, in agreement with experimental results.

To account for the kinetic effect of the electric field, we note that Brandon and Southon~\cite{Brandon1968,Southon1968}  independently connected the gas kinetics theory with the high-electric field, allowing us to approximate the time for the field-ion imaging gas molecules (i.e., H$_2$ molecules) adsorption on the electrically charged surface of the APT specimen. 
The total adsorption of gas per unit area per time, $\Phi$ ($\frac{\rm H{_2}\,molecules}{\rm m^2\cdot sec}$) can be obtained by combining the equation of the classical gas kinetics factor (see Eq.~\ref{kinetic_eq2}), electric-field enhancement factor (see Eq.~\ref{kinetic_eq3}), and probability, $P$, that a gas molecule is ionized on its way to the tip.

\begin{equation}
\Phi = \Phi_{0}\zeta(1-P) \quad ,
\label{kinetic_eq1}
\end{equation}
where $\Phi_{0}$ is the gas kinetic arrival rate. We assumed $P$ as zero to maximize the net adsorption rate of H. The flux of H that arrives at the surface of the Na nanosized tip in the absence of an electric field ($\mathcal{E}$) is given by

\begin{equation}
\Phi_{0} = \frac{p}{\sqrt{2\pi Mk_{\rm B}T}} \quad ,
\label{kinetic_eq2}
\end{equation}
where $p$ is the partial pressure of H$_2$, $M$ is the molecular mass of H$_2$ [2 atomic mass unit (amu)], and $T$ is the base temperature of the specimen (i.e., 30\,K). Since H is the predominant residual gas in metal vacuum chambers at low pressures, here we assume $p$ equal to the analysis chamber pressure during the APT measurement that is ${3.43\times10^{-14}}$\,bar. When an electric field is applied to the Na tip, the H supply is enhanced by a factor of $\zeta$:

\begin{equation}
\zeta = \frac{\mu^{\rm H_2}_{\rm dipole}\mathcal{E}}{k_{\rm B}T} + \sqrt{\frac{\pi\alpha\mathcal{E}^2}{2k_{\rm B}T}}
{\rm erf}\left[\left( \frac{\alpha\mathcal{E}^2}{2k_{\rm B}T}\right)^2\right] \quad ,
\label{kinetic_eq3}
\end{equation}
where $\mu^{\rm H_2}_{\rm dipole}$ is the permanent dipole moment of gas for a H$_2$ molecule and $\alpha$ is its polarizability. Assuming that $\mu^{\rm H_2}_{\rm dipole}$ is zero and the error function gives a value of 1, the equation simplifies to Eq.~\ref{kinetic_eq4}:
\begin{equation}
\zeta \approx
\sqrt{\frac{\pi\alpha\mathcal{E}^{2}}{2k_{\rm B}T}} \quad .
\label{kinetic_eq4}
\end{equation}

The following experimental and theoretical values used in our calculation, $\alpha_{\rm H_2} = 5.314$ [atomic unit (au)]~\cite{OLNEY199759} and assuming  an electric field of $\mathcal{E} = 2$\,V/{\AA} for Na, yield the total supply of gas $\Phi_{0} \approx 48$\,$\frac{\rm H{_2}\,molecules}{\rm m^2\cdot sec}$ and $\zeta = 11.5$, which results in $\Phi = 549$\,$\frac{\rm H{_2}\,molecules}{\rm m^2\cdot sec}$ striking the Na tip. Assuming that an area of an APT specimen is ${\pi\cdot(100\times100)}$\,nm$^2$, the approximate time to achieve a monolayer coverage of H molecules on the Na tip is over 1800\,years. This implies that the hydrogen detected over the course of an experiment is mostly supplied from chemisorbed or physisorbed hydrogen on the specimen's shank that surface-diffuses towards the apex, and has little to do with the direct chemisorption or ionisation of hydrogen at the apex itself. The field-free adsorption properties of the surface are hence likely the most critical parameter to consider. 

\subsection{Summary}
Our discussions consistently suggest that the possible interaction between the residual H$_2$ gas in the UHV chamber and an APT specimen is both thermodynamically and kinetically almost negligible, which agrees with previous reports~\cite{doi:10.1021/nn305029b,doi:10.1017/S143192761500032X}. As illustrated in Fig.~\ref{fig5}(f), this supports the other possible origin of the H$_2$ detection - out-gassing and diffusion of H from the specimen itself [red-solid arrows in Fig.~\ref{fig5}(f)], with H introduced during the preparation, and/or the specimen holders (i.e., Cu holder and Si coupon)~\cite{BREEN2020108,doi:10.1063/1.1597372} that were exposed to air before going into the UHV system. Subsequent surface diffusion indicated by blue arrows in Fig.~\ref{fig5}(f), leads to a migration of the H at the surface towards the APT specimen apex to finally get desorbed and ionized as shown by the purple arrow in Fig.~\ref{fig5}(f)~\cite{ANTCZAK200739}. Both processes (i.e., ad-/absorption) strongly depend on the material's properties.

\section{\label{Sec: Conclusions} Conclusions}
In conclusion, we studied the origin of residual hydrogen in APT experiments by combining APT analyses of Na and Pt metals and DFT calculations regarding the thermodynamic stability of H-covered metal surfaces (Na, K, Pt, and Pd). In our APT measurements, large peaks for H residuals are observed for Pt, as commonly observed in most APT experiments, whereas no or negligible amounts of residual H are measured in Na. The surface phase diagrams of H-exposed metal surfaces constructed based on DFT calculations indicate that H contamination can easily occur for Pt and Pd surfaces, but not for Na and K surfaces, at least within the conditions of our experiments. This combined result provides the insight that residual hydrogen in APT measurements mostly originates from H contamination of materials during the specimen preparation and transport and not from the background H$_2$ gas alone. The design of novel instrumentation remains very important. However, careful specimen preparation and transport, involving cryogenic workflows and vacuum transfer, seem to be lower hanging fruits that can lead to substantial improvements in data quality. The possible coatings of specimens and holders with materials on which hydrogen adsorption is not favourable should also be explored in the future. 

\section{acknowledgments}
Funding by the German Research Foundation (Deutsche Forschungsgemeinschaft (DE)) within the framework of SFB1394, project number 409476157 is gratefully acknowledged. S.-H.K.and B.G. acknowledge financial support from the ERC-CoG-SHINE771602. S.-H.K.and B.G. also acknowledge Uwe Tezins, Christian Broß, and Andreas Sturm for their support to the FIB and APT facilities at MPIE. S.-H.K.and B.G. are grateful for the Max-Planck Society and the BMBF for the funding of the Laplace and the UGSLIT projects respectively, for both instrumentation and personnel. 

\section{Data availability statement}
All DFT calculated data used in this work is available in the Pyiron repository and can be given access to upon request.

\section{Supporting Information}
Supporting Information is available.

\bibliography{ms}

\end{document}


\title{\underline{Supplemental Material}\\ Origins of the hydrogen signal in atom probe tomography}

\author{Su-Hyun \surname{Yoo}}
\email{yoo@mpie.de}
\affiliation{Department of Computational Materials Design, Max-Planck-Institut f\"{u}r Eisenforschung GmbH, Max-Planck-Str. 1, D-40237 D\"{u}sseldorf, Germany}
\author{Se-Ho \surname{Kim}}
\affiliation{Department of Microstructure Physics and Alloy Design, Max-Planck-Institut f\"{u}r Eisenforschung GmbH, Max-Planck-Str. 1, D-40237 D\"{u}sseldorf, Germany}
\author{Eric \surname{Woods}}
\affiliation{Department of Microstructure Physics and Alloy Design, Max-Planck-Institut f\"{u}r Eisenforschung GmbH, Max-Planck-Str. 1, D-40237 D\"{u}sseldorf, Germany}
\author{Baptiste \surname{Gault}}
\email{b.gault@mpie.de}
\affiliation{Department of Microstructure Physics and Alloy Design, Max-Planck-Institut f\"{u}r Eisenforschung GmbH, Max-Planck-Str. 1, D-40237 D\"{u}sseldorf, Germany}
\affiliation{Department of Materials, Royal School of Mines, Imperial College, London, SW7 2AZ, United Kingdom}
\author{Mira \surname{Todorova}}
\affiliation{Department of Computational Materials Design, Max-Planck-Institut f\"{u}r Eisenforschung GmbH, Max-Planck-Str. 1, D-40237 D\"{u}sseldorf, Germany}
\author{J\"{o}rg \surname{Neugebauer}}
\affiliation{Department of Computational Materials Design, Max-Planck-Institut f\"{u}r Eisenforschung GmbH, Max-Planck-Str. 1, D-40237 D\"{u}sseldorf, Germany}

\date{\today}

\maketitle


\begin{table*}[!htbp]
\caption{Computational details for metal slab calculations. Used exchange-correlation approximation functional, size of supercell, $k$-point mesh, the number of valence electrons for the metal atom, thickness of vacuum, thickness and type [e.g., symmetric (asymmetric) slab is abbreviated to Symm. (Asymm.)] of slab, and the numbers of relaxed and fixed layers are listed.}
\label{s_tab1}
\begin{ruledtabular}
\begin{tabular}{c cccc}
\noalign{\vskip 2mm}
System & Na(110) & K(110) & Pd(111) & Pt(111) \\
\noalign{\vskip 1mm}
\hline
\noalign{\vskip 2mm}
Functional & PBE~\cite{Perdew1996,Perdew1997} & PBE~\cite{Perdew1996,Perdew1997} & PBE~\cite{Perdew1996,Perdew1997} & RPBE~\cite{Perdew1996,Hammer1999} \\
Size of supercell & $p(3\times3)$ & $p(3\times3)$ & $p(2\times2)$ & $(3\times2\sqrt{3})$rect \\
$k$-point mesh & $\Gamma$ $4\times4\times1$ & $\Gamma$ $4\times4\times1$ & $\Gamma$ $4\times4\times1$ & $\Gamma$ $8\times6\times1$ \\
Valence & 7 & 7 & 16 & 10 \\
Vacuum thickness (${\rm \AA}$) & 18 & 18 & 18 & 12 \\
Slab thickness (atomic layer, AL) & Asymm., 8 & Asymm., 8 & Symm., 13 & Asymm., 4 \\
Relaxed layers (AL) & 5 & 5 & 6 & 2 \\ 
Fixed layers (AL) & 3 & 3 & 7 & 2 \\
\noalign{\vskip 2mm}
\end{tabular}
\end{ruledtabular}
\end{table*}
\FloatBarrier

\begin{figure}[!htbp]
\center
\includegraphics[width=0.9\columnwidth]{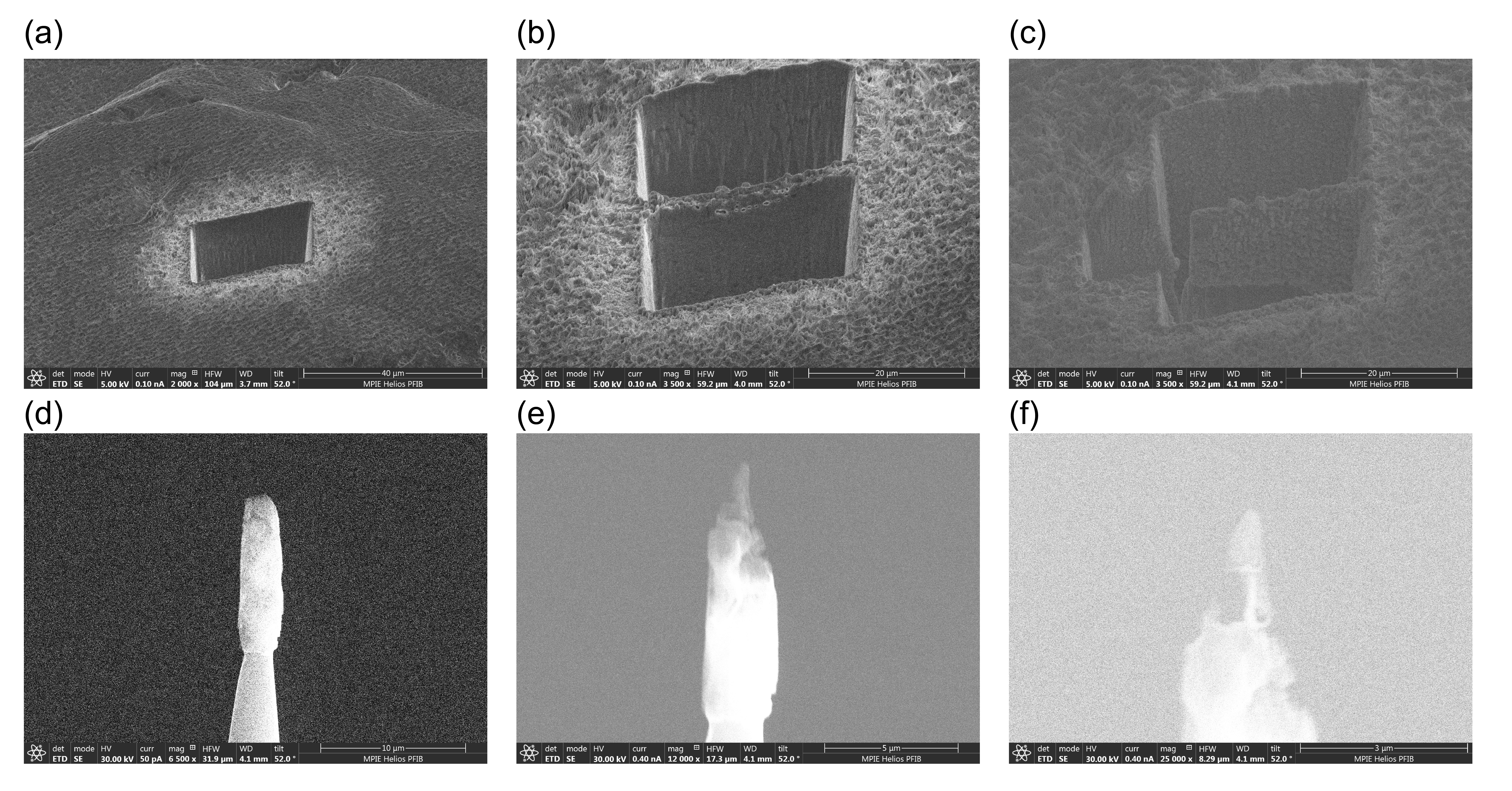}
\caption{Uncontrollable melting of the Na sample at room temperature. (a-c) Sectioning process. (d-f) Annular milling process.}
\label{S_melting_Na}
\end{figure}
\FloatBarrier

\begin{figure}[!htbp]
\center
\includegraphics[width=0.7\columnwidth]{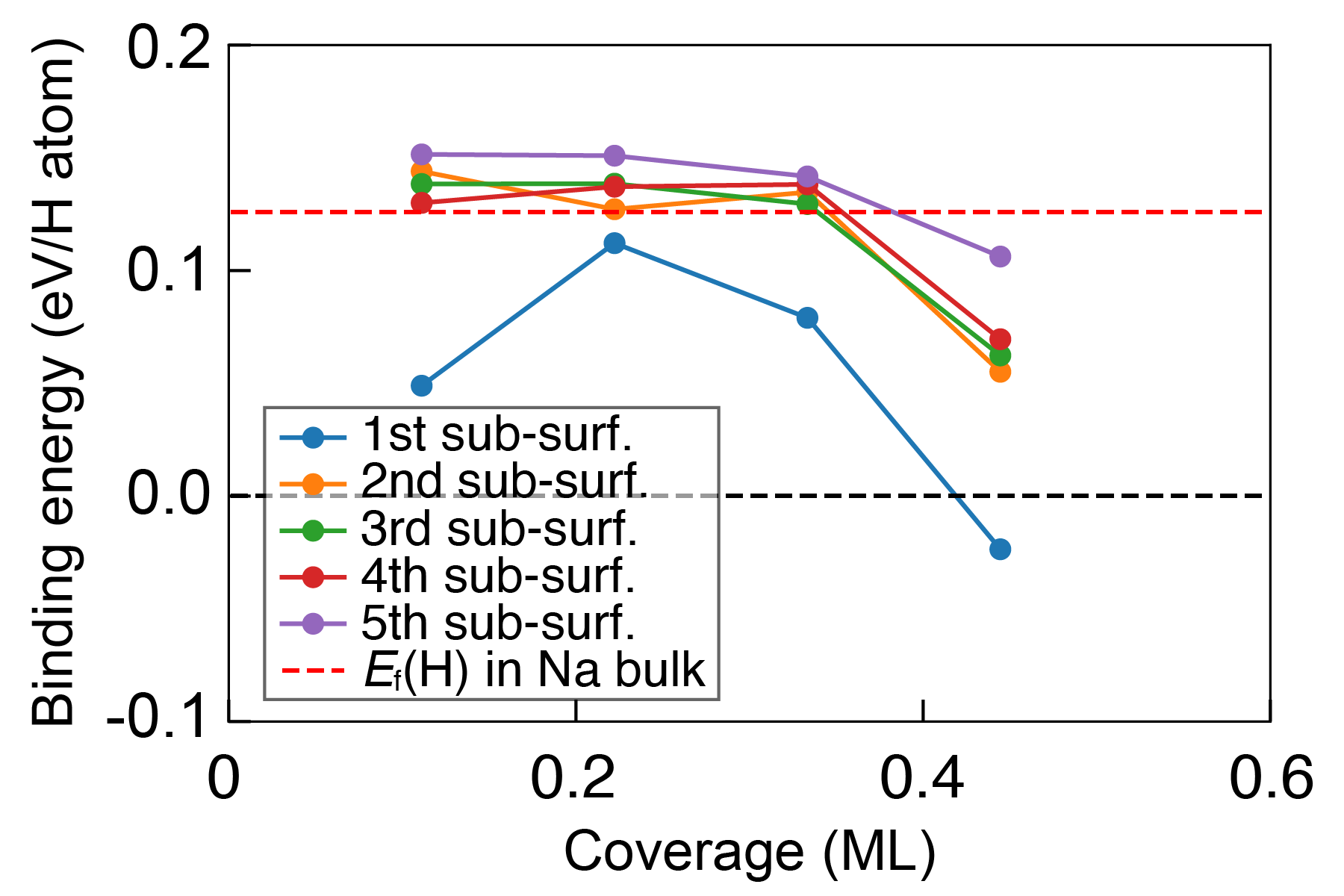}
\caption{Calculated binding energies of a H interstitial placed in a tetrahedral site in the 1st (blue), the 2nd (orange), the 3rd (green), the 4th (red), and the 5th (purple) subsurface layer are plotted as a function of H coverage. The formation energy of a H interstitial in a tetrahedral site in a Na $p({4\times4\times4})$ bulk supercell is indicated by a horizontal red dashed line.}
\label{S_Eb_Hsub}
\end{figure}
\FloatBarrier

\begin{figure}[!htbp]
\center
\includegraphics[width=0.9\columnwidth]{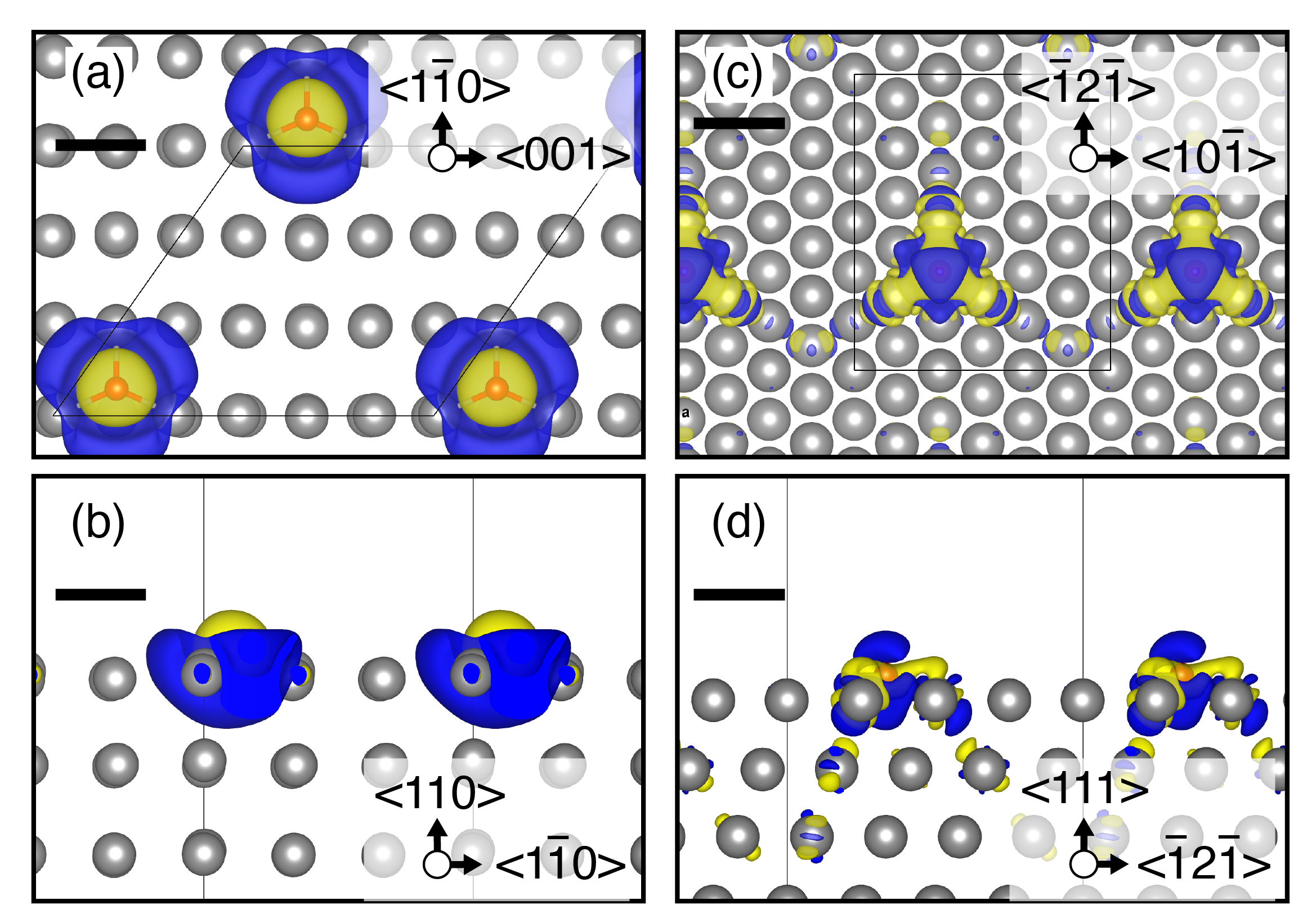}
\caption{Electron density differences ($\Delta \rho = \rho_{\rm H-surf} - \rho_{\rm surf} - \rho_{\rm H}$) showing the redistribution of electron density upon adsorption of H on the metal surfaces [(a)$\backsim$(b) 0.11\,ML H-Na(110) and (c)$\backsim$(d) 0.08\,ML H-Pt(111), respectively]. The top (bottom) panels are top- (side-) view of 3D electron density differences. Yellow (blue) areas correspond to electron accumulation (depletion). Scale bar is 3\,${\rm \AA}$.}
\label{S_cdd_3d}
\end{figure}
\FloatBarrier

\begin{figure}[!htbp]
\center
\includegraphics[width=0.8\columnwidth]{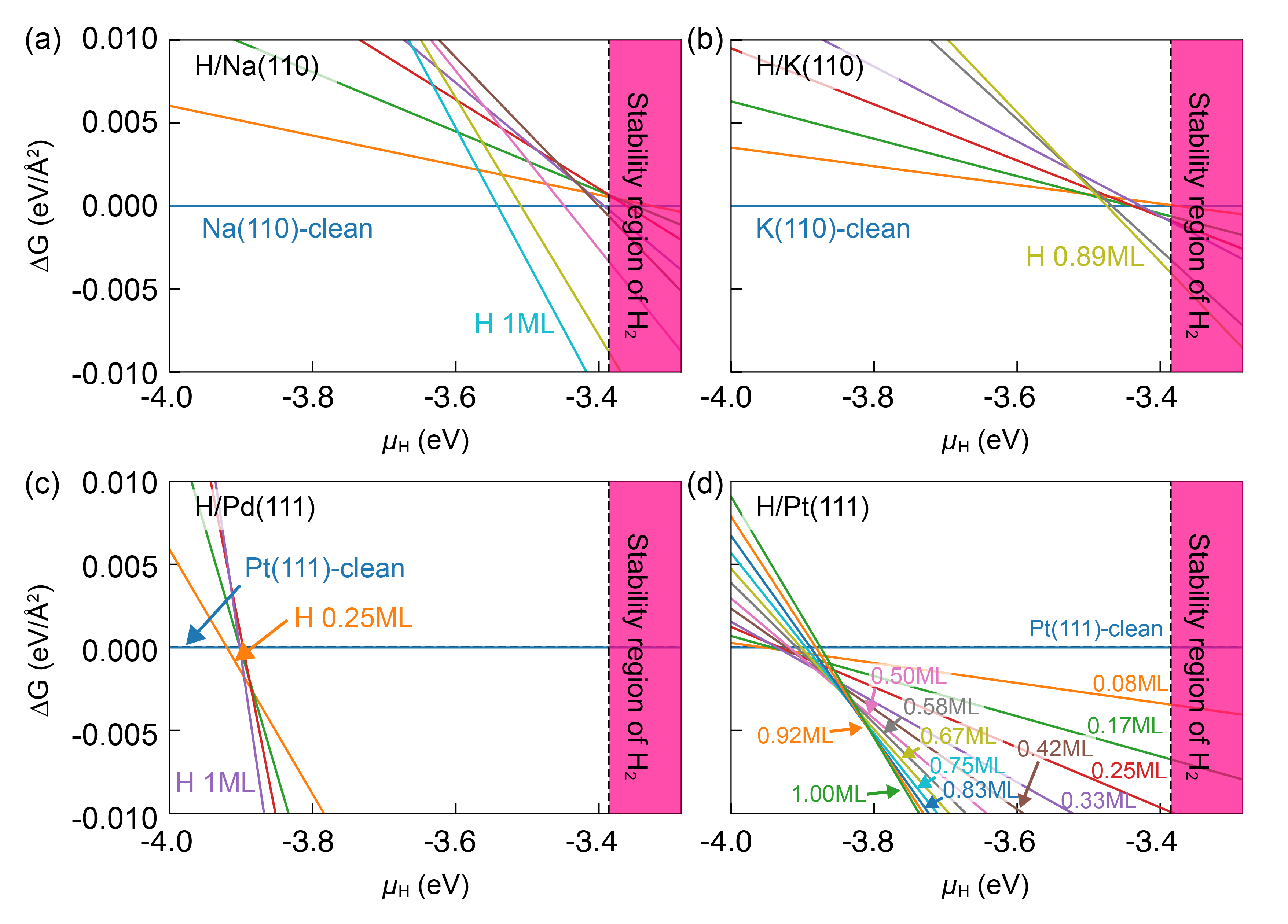}
\caption{The Gibbs free energy differences of surface phases with respect to the H-free clean surface are plotted as a function of the H chemical potential for (a) Na(110), (b) K(110), (c) Pd(111), and (d) Pt(111), respectively. Each color line indicates each surface phase (i.e., with different H coverage). Purple color area indicates the stablility region for H$_2$ molecules.}
\label{S_dG}
\end{figure}
\FloatBarrier

\begin{figure}[!htbp]
\center
\includegraphics[width=0.5\columnwidth]{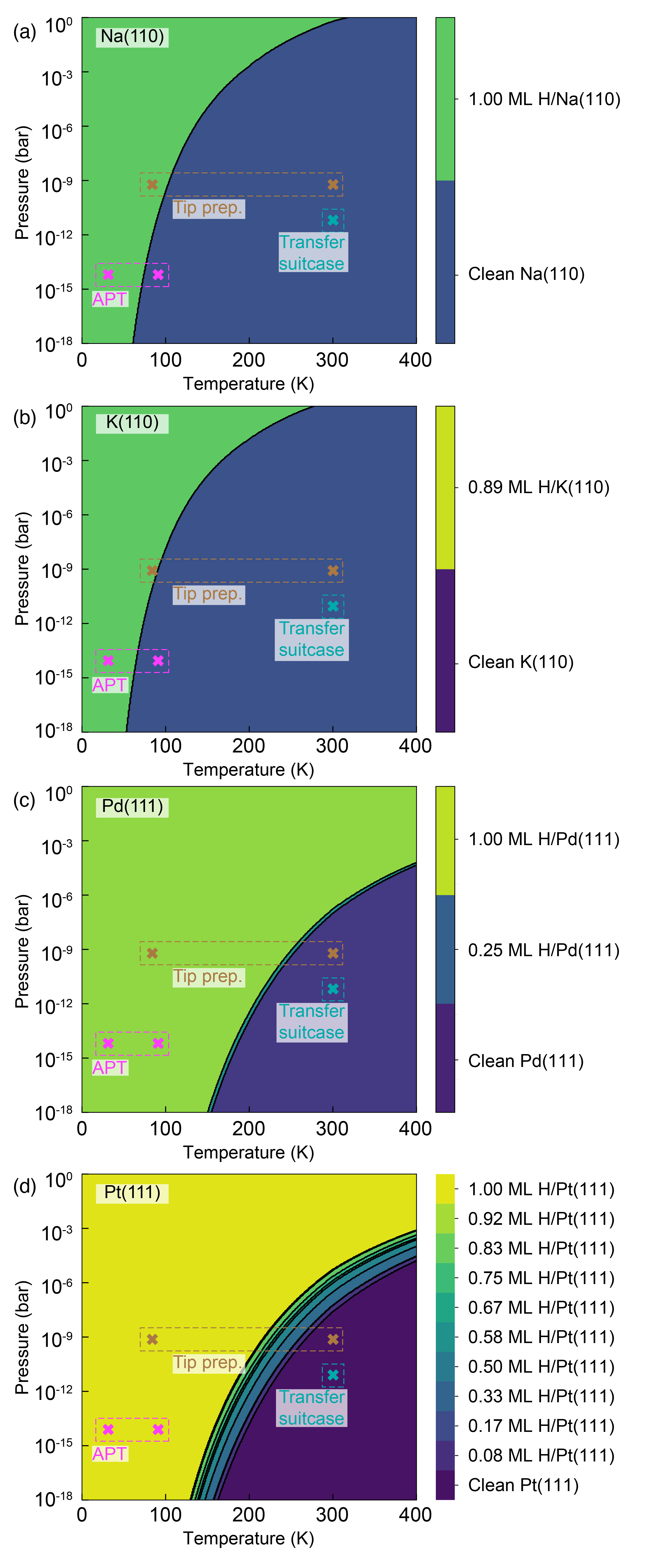}
\caption{Surface phase diagrams for metal surfaces [(a) Na(110), (b) K(110), (c) Pd(111), and (d) Pt(111)] in equilibrium with a surrounding H$_2$ gas as functions of temperature and pressure. Each coloured region indicates the most favorable surface phase. Pink, brown, and cyan crosses show the experimental conditions used in this work for APT measurements, tip preparation, and tip transportation, respectively.}
\label{S_phase_diagram_inv_metals}
\end{figure}
\FloatBarrier

\newpage
\bibliography{supplement}